\documentclass [12 pt]{article}
\usepackage{psfig}
\usepackage{epsfig}
\begin{document}
\title {Electric fields of a uniformly charged elliptical beam.}
\author{George Parzen}
\date{August 3, 2001 \\BNL/SNS Technical note \\NO.098 }
\maketitle
\begin{abstract}
This paper presents results for the electric field due 
to  a  uniformly charged elliptical beam in the region 
outside the beam.
\end{abstract}

\section*{Introduction}

	This paper presents results for the electric field due to  a  uniformly
charged elliptical beam outside the beam. Results for the field inside inside the beam are well known \cite{Kellog53,Teng59} The beam being considered extends 
indefinitly in the $z$ direction and has an elliptical boundary in  $x$ and 
$y$ given by
\begin{equation} 
 x^2/a^2+y^2/b^2=1
\end{equation}
The charge density, $\rho (x,y,z)$ is uniform within the elliptical boundary,
zero outside the elliptical boundary, and does not depend on $z$.  The results 
given below depend on the observation made by B. Houssais \cite{Sacherer71},
that the result for the electric field of a gaussian charge distribution given
by W. Kellog \cite{Kellog53} as a one dimensional integral would hold for any 
elliptical charge distribution as defined below. This may be stated as follows.
Let the charge distribution be given as
\begin{equation} 
\rho (x,y,z)=\lambda n(x,y)             
\end{equation}
where $\lambda$ is the charge per unit length and
\begin {equation} 
\int dxdy \; n(x,y)=1
\label{3} 
\end {equation}
	A charge distribution will be called elliptical if $n(x,y)$ can be 
written as
\begin{eqnarray}
n(x,y) &=& \hat n(T)/\pi ab  \nonumber\\
T &=&  x^2/a^2+y^2/b^2       \label{4}
\end{eqnarray}
For the uniform elliptical beam , $\hat n(T)$ is given by
\begin{eqnarray}
\hat n(T) &=& 1, \;\; T\leq 1  \nonumber\\
\hat n(T) &=& 0, \;\; T> 1
\end{eqnarray}
For a Gaussian beam, $\hat n(T)$ is given by
\begin{eqnarray}
\hat n(T) &=& exp(-T)
\end{eqnarray}
One can show, using Eq.~\ref{3}, that $\hat n(T)$ obeys the equation
\begin {equation} 
\int_{0}^{\infty} dT \; \hat n(T) = 1
\end {equation}
The generalization of the Kellog result for any elliptical beam is then
\begin{eqnarray}
E_x &=& 2\lambda \int_{0}^{\infty} dt \frac{\hat n(\hat T)}{(a^2+t)^{3/2}(b^2+t)^{1/2}}                            \nonumber\\
\hat T &=&  x^2/(a^2+t)+y^2/(b^2+t)     \label{7}
\end{eqnarray}
A similar result, with $a,b$ and $x,y$ interchanged will give $E_y$

\section*{Electric fields for x,y inside the beam}

As a first step, the fields inside a uniformly charged  elliptical beam  will 
be found using Eq.~\ref{7} .In  this case, $\hat T$ is always $\leq 1$ since
for $t=0$ , $\hat T=x^2/a^2+y^2/b^2$,  which is $\leq 1$ for x,y inside the beam,
and decreases further for larger t. Eq.~\ref{7} then becomes
\begin{eqnarray}
E_x &=& 2\lambda x  \int_{0}^{\infty} dt \frac{1}{(a^2+t)^{3/2}(b^2+t)^{1/2}}                            \nonumber\\
\hat T &=&  x^2/(a^2+t)+y^2/(b^2+t)     \label{8}
\end{eqnarray}
The integral in Eq.~\ref{8} can be done using the result
\begin{eqnarray}
\int_{t_1}^{\infty} dt \frac{1}{(a^2+t)^{3/2}(b^2+t)^{1/2}} &=& 
      2 \frac{1}{(a^2+t_1)^{1/2}} \frac{1}{(a^2+t_1)^{1/2}+(b^2+t_1)^{1/2}}
\label{9}         
\end{eqnarray}
This gives 
\begin{eqnarray}
E_x &=& 4\lambda x \frac{1}{a(a+b)}
\end{eqnarray}
and a similar result for $E_y$ with a and b interchanged
and x replaced by $y$.

\section*{Electric fields outside the beam when $y=0$}

As the next step, the fields outside  a uniformly charged  elliptical beam  will 
be found using Eq.~\ref{7} for the case when $y=0$. The results in this case are
simpler and the mathematics is easier to comprehend. In  this case, $\hat T$ is 
$>1$ for $t=0$ since 
for $t=0$ , $\hat T=x^2/a^2+y^2/b^2$,  which is $> 1$ for x,y outside the beam.
For larger $t$, $\hat T$ decreases and reaches the vaue of 1 at $t=t_1$, 
and at still larger $t$, $\hat T$ decreases further always remaining 
smaller than 1.The integral in Eq.~\ref{7} then goes from $t=t_1$, to
$t=\infty$. Eq.~\ref{7} then becomes
\begin{eqnarray}
E_x &=& 2\lambda x  \int_{t_1}^{\infty} dt \frac{1}{(a^2+t)^{3/2}(b^2+t)^{1/2}}                            \nonumber\\
\hat T &=&  x^2/(a^2+t)     \label{11}  \\
t_1 &=& x^2-a^2   \nonumber  \\
y &=& 0
\end{eqnarray}
Using Eq.~\ref{9}. one finds
\begin{eqnarray}
E_x &=& 4\lambda \frac{1}{x+(x^2+b^2-a^2)^{1/2}}  \nonumber \\
E_y &=& 0       \\
y &=& 0    \nonumber
\end{eqnarray}
$E_{xx}=\partial E_x/\partial x$ is given by
\begin{eqnarray}
E_{xx} &=& -\frac {E_x}{(x^2+b^2-a^2)^{1/2}} 
\end{eqnarray}

\section*{Electric fields outside the beam when $y\neq 0$}

As the final step, the fields outside  a uniformly 
charged  elliptical beam  will 
be found using Eq.~\ref{7} for the general case.  
In  this case, $\hat T$ is 
$>1$ for $t=0$ since 
for $t=0$ , $\hat T=x^2/a^2+y^2/b^2$,  which is $> 1$ 
for x,y outside the beam.
For larger $t$, $\hat T$ decreases and reaches the vaue of 1 at $t=t_1$, 
and at still larger $t$, $\hat T$ decreases further always remaining 
smaller than 1.The integral in Eq.~\ref{7} then goes from $t=t_1$, to
$t=\infty$. Eq.~\ref{7} then becomes
\begin{eqnarray}
E_x &=& 2\lambda x  \int_{t_1}^{\infty} dt \frac{1}{(a^2+t)^{3/2}(b^2+t)^{1/2}}            \nonumber \\
x^2/(a^2+t_1)+y^2/(b^2+t_1) &=& 1 
\label{15}       
\end{eqnarray}
$t_1$ is the positive root of the equation
\begin{eqnarray}
x^2/(a^2+t_1)+y^2/(b^2+t_1) &=& 1 
\label{16}       
\end{eqnarray}
The quadratic equation for $t_1$, Eq.~\ref{16}, can be solved 
to give
\begin{eqnarray}
t_1 &=& (B^2/4+C)^{1/2}+B/2   \nonumber  \\
B &=& x^2+y^2-a^2-b^2  \label {17}    \\
C &=& x^2 b^2 + y^2 a^2-a^2 b^2  \nonumber 
\end{eqnarray}
Eq.~\ref{15} gives the result for $E_x$
\begin{eqnarray}
E_x &=& 4 \lambda x \frac{1}{(a^2+t_1)^{1/2}} \frac{1}{(a^2+t_1)^{1/2}+(b^2+t_1)^{1/2}}  
\end{eqnarray}
and a similar result for $E_y$ with a and b interchanged
and $x$ replaced by $y$.

It may be usefull to  also have results for the derivatives
of the fields, $E_{xx}, E_{yy}, E_{xy}=E_{yx}$, where
$E_{xx}=\partial E_x/\partial x$, $E_{yy}=\partial E_y/\partial y$ and $E_{xy}=\partial E_x/\partial y$. $E_{xx}$ is found using
 Eq.~\ref{15} for $E_x$
\begin{eqnarray}
E_{xx} &=& \frac{E_x}{x}- 2 \lambda x \frac{1}{(a^2+t_1)^{3/2}(b^2+t_1)^{1/2}} \frac{dt_1}{dx}
\end{eqnarray}
$dt_1/dx$ can be found from Eq.~\ref{16} for $t_1$ as
\begin{eqnarray}
\frac{dt_1}{dx} &=& 2 x \frac {(a^2+t_1) (b^2+t_1)^2}{x^2(b^2+t_1)^2+
    y^2(a^2+t_1)^2}
\end{eqnarray}
This gives for $E_{xx}$
\begin{eqnarray}
E_{xx} &=& \frac{E_x}{x}- 4 \lambda x^2 
\frac {(a^2+t_1)^{-1/2}(b^2+t_1)^{3/2}}{x^2(b^2+t_1)^2+y^2(a^2+t_1)^2}
\end{eqnarray}
$E_{yy}$ and $dt_1/dy$ can be found by interchanging x and y,
and a and b.
$E_{xy}$ can be found in  the same way as
\begin{eqnarray}
E_{xy} &=& - 4 \lambda x y 
\frac {(a^2+t_1)^{1/2}(b^2+t_1)^{1/2}}{x^2(b^2+t_1)^2+y^2(a^2+t_1)^2}
\end{eqnarray}
%

\end{document}